\documentstyle[12pt]{article}
\title{NEUTRINO ASTRONOMY:\\
Physics Goals, Detector Parameters\thanks{Talk given at
the OECD Megascience Forum Workshop, Taormina, Italy, 22/23 May 1997.  Work
supported in part by the U.S. Department of Energy.}}
\author{T.K. Gaisser\\
Bartol Research Institute, University of Delaware\\
Newark, DE 19716 USA}
\begin{document}
\maketitle
\section{Introduction}

Observing astrophysical neutrinos opens a new window on
the Universe.  The claim is already justified
by measurement of solar neutrinos and by Supernova 1987A.
Several new detectors are in operation or under construction
to study in detail the energy range from a few MeV to tens of GeV.
The primary focus of a kilometer-scale detector will be to observe
neutrinos at higher energy.

After reviewing the energetics of extragalactic cosmic rays,
which are closely related to energetic neutrinos in most
scenarios, I list some possible sources that have been suggested
and describe their neutrino signatures.
Next I discuss the
nature of the neutrino signals and how they can be detected.
The summary section includes some remarks on the prospects
for actually finding signals of high energy astrophysical
neutrinos.  I also comment on the backgrounds of atmospheric
neutrinos, which may also serve as a calibration source.

Apart from a brief mention of diffuse galactic neutrinos, I
focus in this talk on neutrinos associated with the highest
energy cosmic rays, which are assumed to be of extragalactic origin.
I also do not address here the important subject of neutrinos from
decay of neutralinos, which was covered in David Schramm's talk.
(See also Ref. \cite{Bottino} for a recent treatment of the subject.)

\section{Extragalactic cosmic radiation}

High energy neutrinos require
the presence of energetic hadrons.  It is therefore of interest
to use the observed energy density in cosmic rays to
estimate the level of neutrinos that might be expected.
I discuss the extragalactic component of the cosmic radiation
and corresponding extragalactic sources of neutrinos.
An analogous argument can be made for sources inside the
galaxy \cite{GHS}.

Fig. 1\footnote{From Ref. \cite{PP} with the addition of
data from Tibet \cite{Tibet}.} shows the cosmic-ray spectrum with an assumed
extragalactic component fitted above the ``ankle'' and a
GZK cutoff around $5\times 10^{19}$~eV.  The energy content
of this component is $\sim 3\times 10^{-19}$~erg~cm$^{-3}$,
assuming an extrapolation to low energy with an $E^{-2}$
differential spectrum.  (The total energy content would be higher
if the spectrum of the extragalactic component below the ankle
is steeper.)  The power required to generate this energy
density in $10^{10}$~yrs is $\sim 3\times 10^{37}$~erg/s per (Mpc)$^3$.
This works out to
\begin{itemize}
\item $\sim3\times10^{39}$ erg/s per galaxy,
\item $\sim3\times10^{42}$ erg/s per cluster of galaxies,
\item $\sim2\times10^{44}$ erg/s per active galaxy, or
\item $\sim2\times10^{52}$ erg per cosmological Gamma Ray Burst (GRB).
\end{itemize}
The coincidence between these numbers and the observed
energy output of these objects in other wavelengths, especially in the
last two cases, is the basis for their use as models of
origin of the highest energy cosmic rays.

Suppose now that there is the same energy density in neutrinos,
$\rho_E\,\sim\,3\times 10^{-19}$~erg/cm$^3$
with a spectrum\begin{equation}
E_\nu{dN\over dE_\nu}\;=\;A\,E^{-\gamma}\;cm^{-2}\,s^{-1}\,sr^{-1}
\end{equation}
that continues to a maximum neutrino energy, $E_{max}$.
Then \begin{equation}
\int A\,E_\nu^{-\gamma}\,dE_\nu\;=\;{c\over 4\pi}\,\rho_E
\end{equation}
normalizes the (diffuse) neutrino flux.
The resulting signal depends on the neutrino spectrum.
For $\gamma = 1$ and $E_{max} \sim 10^8$~GeV the rate
of neutrino-induced muons is about 50 per km$^2$ per year
and depends only weakly on $E_{max}$.  For $\gamma=0$
up to the same $E_{max}$ the corresponding number
is 20 events per year, with the rate nearly inversely
proportional to $E_{max}$.

\section{Possible sources and their neutrino signatures}
Is there a physical realization of a source that would
produce approximately equal energy densities in cosmic-rays
and neutrinos?  Consider a generic source
in which protons and electrons are accelerated to high
energy at shocks.  The acceleration process is fast but the charged particles
remain trapped in the diffuse, turbulent magnetized
plasma until they radiate photons, the electrons by
bremsstrahlung and the protons by photoproduction.  Neutrons
from $p+\gamma\rightarrow n +\pi^0$ escape to become cosmic-rays.
Neutrinos from charged pions will have an energy comparable
to the energy in escaping cosmic-rays.  The neutrino/$\gamma$-ray
ratio, among other things, reflects the ratio of accelerated
electrons to protons in the source.  This is a crude generic
description of some models of AGN jets \cite{Mannheim,rprother}.

Many of the suggested sources of extragalactic cosmic rays 
including AGN \cite{Biermann}, clusters of galaxies 
\cite{Norman,Kang,Dar,Berezinsky}, cosmological gamma-ray bursts \cite{Waxman}
and radiation from topological defects \cite{Sigletal}
also imply a correlated flux of high energy neutrinos.
In addition, there will be neutrinos at some level
from cosmic-ray interactions on the 2.7 $^\circ$K background
radiation \cite{Yoshida1}.

In Fig. 2 and Table 1 I summarize the event rates in
a kilometer-cubed detector from various sources of
high-energy neutrinos.   The figure displays the rates
for neutrino-induced muons and the table includes
events generated by electron-neutrinos as well, including
separately the rate of events at the Glashow resonance.
The diffuse fluxes referred to here are obtained by
integrating over all sources in the Universe for each model
(e.g. all blazars or all GRB, etc.).  A discussion of the
relation between signals from individual point sources
and the corresponding diffuse fluxes will be made elsewhere \cite{GHS2}.
Generally, the signal rate from a nearby point source will be
significantly lower than the corresponding diffuse signal, as
illustrated by the example of 3C273 in Ref. \cite{GHS}.
\begin{table}\caption{Diffuse neutrino rates}
\begin{tabular}{lcccc}\hline\\
Source & $\nu\rightarrow\mu$ & $E_\mu$ (TeV) & $\nu_e+\bar{\nu}_e$
   & \\
   & per km$^2$ yr. &  (median) & (total rate per km$^3$)&
$\bar{\nu}_ee^-$ only\\ \hline
Atmospheric & $5\times 10^5$ & 0.1 &  &  \\
Neutralinos & $<10^4$ & $<0.1$ &  &  \\
Prompt $\mu$ & 400--$\sim10^4$ & 0.4 &  & \\
Galactic center & 300--1000& 0.4 &  & \\
Galaxy clusters & $<5000$ & 1.8 &$<3000$ & $<50$\\
GRB sources & 200 & 7 & 40 & 10 \\
AGN jets & 500 & 15 & 70 & 200 \\
AGN cores & $<3\times 10^4$ & 2 & $<5000$ & $<25$\\
GZK & $<1$ & 500 & & \\
Top. Defects & $<1$ & 5000 & &\\ \hline
\end{tabular}
\end{table}
\section{Neutrino signals}
It is clear from Table 1 and Fig. 2 that some discrimination
on neutrino energy will be needed to see the signals of
high-energy neutrino sources above the local atmospheric
backgrounds.  The rate of neutrino-induced muons is given by
\begin{equation}
{d\,Rate\over d\,\ln E_\nu}\;=\;Area(\theta)\times{dN_\nu\over d\,\ln E_\nu}
\times P_{\nu\rightarrow\mu}(E_\nu,E_{\mu,min}),
\end{equation}
where
\begin{eqnarray}
P_\nu&=&\int_{E[\mu,min]}^{E[\nu]}\,N_A\,{d\sigma_\nu(E_\nu)\over dE_\mu}\,
             d\,E_\mu\\
     &\sim& N_A\,\sigma_\nu(E_\nu)\,Range\left[(1-\langle  y\rangle)E_\nu\right
],
\end{eqnarray}
where $N_A$ is Avogadro's number and
$y$ is the fraction of energy to hadrons in a charged current
interaction of a $\nu_\mu$ (or $\bar{\nu}_\mu$).
The last line of this equation allows one to make
reasonably accurate estimates easily which display the
dependence of the signal on neutrino energy, as in Fig. 2.  To display the
dependence on muon energy it is straightforward to include the
convolution of the differential cross section with the neutrino
range as obtained, for example, from Ref. \cite{LS}.

For results
here I have used the cross sections tabulated by Gandhi et al. \cite{Gandhi}.
The corresponding interaction length of neutrinos is shown in Fig. 3.
Note that the Earth, which is $1.1\times 10^5$~km.w.e.
along a diameter, partially obscures upward neutrinos with
$E_\nu> 100$~TeV.
Fig. 4 is an example of the convolution
of a predicted diffuse neutrino flux \cite{Waxman} with the projected area
(assumed to be 1 km$^2$) times $P_\nu$.  The bottom panel
shows the differential rate integrated over the entire
upward hemisphere with the effect of absorption by the
Earth represented by the shading.  Note that this
effect leads to an angular dependence for neutrinos with
$E> 100$~TeV because the vertically upward neutrinos are
absorbed.

For illustrative purposes, I use a simplified treatment
of the range that displays the essential physical effects of
muon propagation.  The energy-loss rate is
\begin{equation}
{d\,E_\mu\over d\,X}\;=\;-\alpha\,-\,E/\xi,
\label{Eloss}
\end{equation}
with $\epsilon_{critical} = \alpha\xi\sim500$~GeV.
The first term in Eq. \ref{Eloss}, which varies slowly with energy,
represents ionization loss, while the last term represents radiative
losses due to pair production, bremsstrahlung and hadronic interactions.
At the critical energy $\epsilon$ the two energy loss terms are equal.
The solution for the avearage energy is
\begin{equation}
E(X)\;\sim\;(E_0\,+\,\epsilon)\,exp[-X/\xi]\,-\,\epsilon,
\end{equation}
and the range $R$ (neglecting fluctuatons) is obtained by setting $E(R)=0$:
\begin{equation}
R\;\approx \,\xi\,\ln(1\,+\,E_0/\epsilon),
\label{range}
\end{equation}
with $\xi\approx 2$~km~water~equivalent.  (See Ref. \cite{LS} for
a nice discussion of the effect of fluctuations, the effect of
which depends on the neutrino spectrum.  Formulas and tables
for muon energy loss in various media are given in Ref. \cite{Lohmann}.)

The muon range is shown in Fig. 5 as a function of its initial energy.
It is interesting
that for $E_{\mu,0}>0.5$~TeV $R>1$~km.  Thus for TeV
(and higher energy) neutrinos the effective volume for neutrino-induced
muons is larger that the physical volume of the detector even for
a kilometer-cubed detector.  This is the reason that this mode
of detection is of such great interest.

A charged-current interaction of a $\nu_\mu$ (or $\bar{\nu}_\mu$)
can occur anywhere between the full range away from the detector to
just outside (or perhaps inside) the detector volume.
Thus the distance a detected muon has traveled before reaching the detector
has a median
\begin{equation}
R_{1\over 2}\;\approx\;{1\over 2}\,\xi\,\ln(1\,+\,E_0/\epsilon),
\end{equation}
with a flat distribution.  The corresponding distribution of
energies at the detector is therefore very broad with a median
\begin{equation}
E_\mu(median)\; \approx\; \epsilon\,\sqrt{(1\,+\,E_0/\epsilon)}\;-\;\epsilon.
\end{equation}
The right-hand curve in Fig. 5 represents this relation.
For example, the median energy at the detector of a muon
that starts with $4$~TeV is $1$~TeV, while for $200$~TeV
it is $10$~TeV.  For $E_\mu\gg \epsilon$
$E_\mu(median)\approx\sqrt{\epsilon\,E_0}$.

As a specific example, I consider the blazar model of diffuse
fluxes of Protheroe \cite{rprother}.  Fig. 6 shows the
differential rate on linear (upper panel) and logarithmic (lower panel)
scales.  Absorption in the Earth has been included.
The thin solid line is the background of atmospheric
neutrinos exclusive of prompt neutrinos.  The dashed
lines give minimal and maximal estimates of the prompt
neutrino background, which I will discuss further below.
A similar model by Mannheim \cite{Mannheim} produces a similar,
but somewhat lower, signal.

A typical neutrino-induced muon from this spectrum will have
initial energy of $\sim 1000$~TeV, for which the median energy
at the detector is $\sim 20$~TeV.  What will such a muon
look like as it propagates through the detector?  The answer is
shown in Figs. 7 and 8, made from calculations by Todor Stanev.
They show the propagation
history of two individual muons through 1 km of water.
In each case the upper panel is on a linear scale and the lower
is the same event on a logarithmic scale.  What is plotted is
the total number of electrons and positrons along the track.
Each burst generates an air shower with typical length
in water of 5 m.  Since for each subshower
\begin{equation}
E_{shower}\;\approx\; 1\;GeV\times Size\; at\; maximum
\end{equation}
for showers in this energy range, the vertical axis can
also be read as number of GeV deposited per burst.

The examples shown are for
$E_\mu=10$~TeV and $E_\mu=100$~TeV at the detector.  The radiative processes
approximately scale with energy in this region, so we can
summarize the situation by noting that a muon with
$E_\mu\gg$TeV will typically generate a couple of bursts
per kilometer in which it radiates about 10 per cent of its
energy.  This is the characteristic signature that a km$^3$
detector should be sensitive to.

\section{Discussion}
\subsection{Some other sources}
The original estimate of diffuse neutrinos from
cores of radio quiet AGN \cite{Stecker,SzRJP,Szabo} is still of interest.
Though the highest predictions of Szabo and Protheroe \cite{Szabo}
are above the Frejus limit \cite{Frejus} and the possibility of neutrinos from
radio quiet AGN has been overshadowed by the
spectacular results from Whipple, HEGRA etc. \cite{Whipple,HEGRA}
on high energy
gamma-rays from blazars, the model itself could still
be valid.  The upper limit allowed by Frejus is
indicated by the upper dotted line in Fig. 2.

At a lower level, in addition to the diffuse fluxes from AGN jets,
there could be production of neutrinos by cosmic rays interacting
with gas in clusters of galaxies \cite{Dar}.  The upper dashed
curve in Fig. 2 is the signal from maximum allowed flux of 
Ref. \cite{Berezinsky} if the same source
accounts for the diffuse extragalactic $\gamma$-ray fluxes above 100~MeV.
Estimates based on cosmic-ray power of current normal
galaxies and AGN are considerably lower than this maximum.
Achieving this upper limit would require a significant ``bright phase''
of cosmic-ray production at early epochs.

GRB sources could also produce an interesting flux of 
diffuse neutrinos if they are the sources of extragalactic
cosmic rays \cite{Waxman}, as shown by the lowest dotted curve in
Fig. 2 and also in Fig. 4.

Another possibility is our own galactic center.
This is of special interest because of the recent
Egret observation \cite{Egret} that the spectrum from the central
region of the galaxy is harder than would be expected if the
$\gamma$-rays were produced from cosmic-rays with the differential
spectrum of -2.7 observed locally.
The lowest dashed curve in Fig. 2 is normalized to
the Egret data at $\approx 2$~GeV extrapolated to
higher energy with a differential index of 2.5
and assuming a similar spectrum of neutrinos 
(from decay of charged pions).  A more quantitative fit
\cite{Stanev} to the whole spectrum, including bremmstrahlung
photons as well as photons from decay of $\pi^0$s,
gives an even harder spectrum at high energy and
a correspondingly higher neutrino signal.

\subsection{Atmospheric neutrinos: a calibration source}

The idea here is to use the atmospheric neutrinos and
their known angular dependence \cite{Lipari,Agrawal} to calibrate the neutrino
detector.  See Fig. 9 for examples of the angular dependence.
Since high energy diffuse sources may also have a similar
angular dependence (depending on their energy spectra)
due to absorption in the earth, some selection on energy will
be important.  There is a useful discussion of the interplay
between energy threshold and angular dependence of high energy
signals and lower energy background in Ref. \cite{Hill}, which
also discusses the contribution of neutrinos from decay of
charmed particles to the atmospheric background.

\subsection{Charm}

The limiting factor, both for using the atmospheric
neutrinos for calibration and in the search for diffuse
sources is the uncertainty
in the component of prompt neutrinos.
The paper of Thunman et al.
\cite{Thunman}  gives a thorough review
of prompt neutrinos, which come primarily from decay
of charmed hadrons.\footnote{Because of the steep rise of the
structure funcions at small $x$, heavy flavors are always
produced predominantly at threshold in hadronic collisions,
so production of beauty will always be suppressed relative
to charm production even at ultra-high energy.
I thank Paolo Lipari for reminding me of this.}
The work of Volkova et al. \cite{Volkova}
gives another view (larger cross section) of prompt neutrinos.
The lower dashed line in Fig. 6
gives the contribution of prompt neutrinos from charm as
calculated in Ref. \cite{Thunman} assuming only gluon-gluon fusion
to be effective in producing charm.  The upper dashed curve
indicates the level of charm that has been suggested by
others if there is significant charm content in the proton \cite{Volkova}.
(The same two curves are shown in Fig. 2 as a pair of light,
solid curves.)

\subsection{The highest energy neutrinos}
Neutrinos associated with the GZK cutoff and especially neutrinos
from decay of topological defects have extremely high energy,
though the fluxes are low.  Detectors with effective
volumes significantly larger than a kilometer cubed would
be desirable, e.g. radio detection \cite{radio,Price,Jelley} 
or air shower detectors
with gigantic acceptance \cite{Yoshida2,Zasetal}.

\noindent
{\bf Acknowledgments.}  I am grateful for helpful conversations
with Chuck Dermer, George Frichter, Paolo Lipari, Al Mann,
Ray Protheroe, Ina Sarcevic, Todor Stanev and Eli Waxman.
This work is support in part by the U.S. Department of Energy
under Grant No. DE-FG02-91ER40626.

\newpage

\begin{center}
FIGURES
\end{center}
\vspace{0.5cm}

\noindent
Fig. 1.  The high energy cosmic-ray spectrum.  The assumed extragalactic
component is indicated by the solid line above $10^{18}$~eV.
\vspace{0.3cm}

\noindent
Fig. 2.  Differential rate of $\nu$-induced, upward muons per logarithmic 
interval of neutrino energy (per km$^2$-yr).  The fluxes are integrated over
the full $2\pi$ steradians from below the horizon, including the effect
of neutrino absorption in the Earth.\\  
{\bf Heavy solid line}: atmospheric neutrinos
from $\pi$- and K-decay;\\ 
{\bf Thin solid lines} ($<10^7$~GeV): ``miminum'' \cite{Thunman}
and ``maximum'' \cite{Volkova} estimates of prompt neutrinos;\\
{\bf Upper dotted line}:
$\nu$ from AGN cores--maximum allowed by Frejus limit \cite{Frejus}
in calculation of \cite{SzRJP};\\
{\bf Lower dotted line}: $\nu$ from AGN cores 
without confinement \cite{Stecker};\\
{\bf Thin solid line} (peak at $10^6$~GeV) AGN jets from Ref. \cite{rprother};\\
{\bf Thin dashed line}: maximum allowed from clusters of galaxies if this
is source of diffuse (extragalactic) $\gamma$-ray 
background \cite{Berezinsky};\\
{\bf Heavy dashed line}: Galactic plane in the direction of the galactic center
(0.73 sr) \cite{Egret};\\
{\bf Heavy dotted line} (peak at $10^5$~GeV): GRB model of Ref. \cite{Waxman};\\
{\bf Upper thin solid line} ($>10^7$~GeV): neutrinos from topological defects
\cite{Sigletal};\\
{\bf Lower thin solid line} ($>10^7$~GeV): example of neutrinos from
interactions of cosmic rays with the microwave background radiation
(maximum red shift = 2; bright-phase index = 2) \cite{Yoshida1}.
\vspace{0.3cm}

\noindent
Fig. 3. Neutrino interaction length vs. energy (from Ref. \cite{Gandhi}).
\vspace{0.3cm}

\noindent
Fig. 4. Convolution of the predicted GRB flux of $\nu_\mu+\bar{\nu}_\mu$
\cite{Waxman} with $P_\nu \times$~detector area (upper panel, arbitrary units) 
and the corresponding response curve (lower panel), 
which shows the relative contribution of neutrinos of different energies
to the predicted signal of upward muons.
The dashed curve in the lower panel shows
the signal when absorption of high-energy neutrinos in the Earth is
neglected, while the solid curve includes the effect of absorption.
\vspace{0.3cm}

\noindent 
Fig. 5. Muon range vs. energy (left solid curve).  The right curve
illustrates the relation between muon energy at production and at
the detector (see text).
\vspace{0.3cm}

\noindent
Fig. 6. Differential rate of neutrino-induced muons per logarithmic
interval of parent neutrino energy for the AGN jet calculation of
Ref. \cite{rprother}.  The thin solid line is the atmospheric
background from decay of pions and kaons and the dashed lines are
minimal \cite{Thunman} and maximal \cite{Volkova}
estimates of prompt neutrinos.  The same information is shown in
both panels (upper with linear scale and lower with logarithmic).
\vspace{0.3cm}

\noindent
Fig. 7. History of a single 10 TeV muon propagating through
one kilometer of water.  Scale can be read as GeV per burst (see text).
Upper and lower panels show the history of the same muon on 
linear and logarithmic scales respectively.
\vspace{0.3cm}

\noindent
Fig. 8. Same as Fig. 6 for a 100 TeV muon.
\vspace{0.3cm}

\noindent
Fig. 9. Angular dependence of upward, $\nu_\mu$-induced muons.
The same information is shown on linear and logarithmic scales
in the upper and lower panels.  In each case the
upper curve is for $E_\mu > 1$~TeV and the lower for $E_\mu > 3$~TeV.
The scale is events per year per bin of $\Delta\cos\theta=0.1$
in a projected area of $2\times 10^4$~m$^2$.
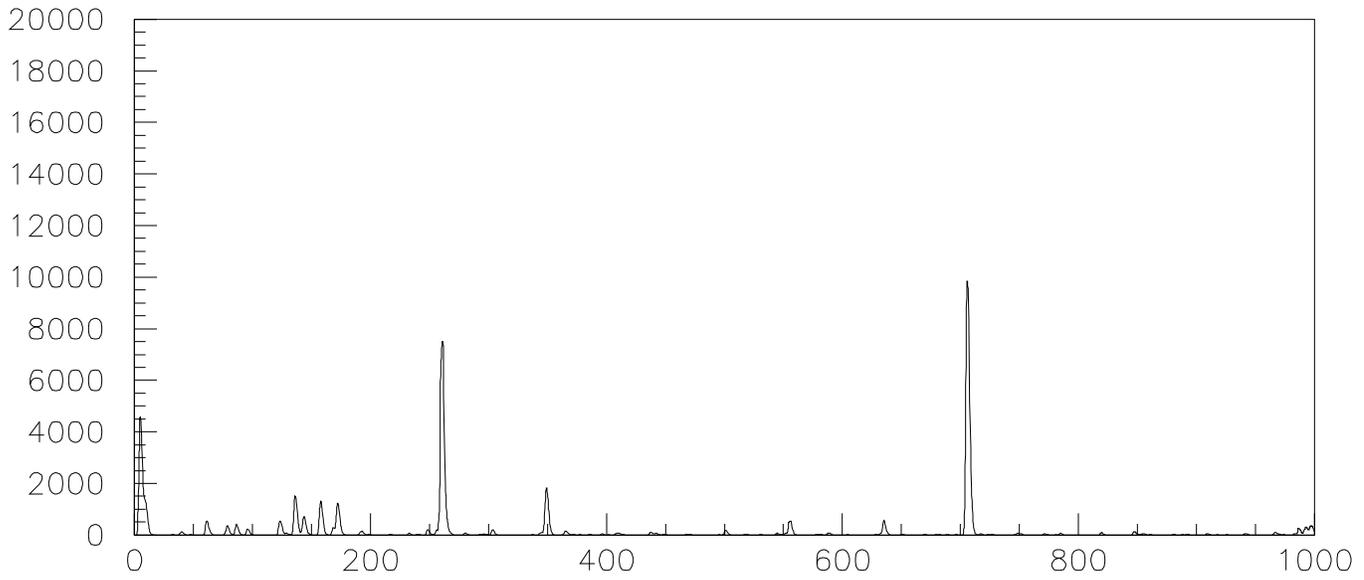
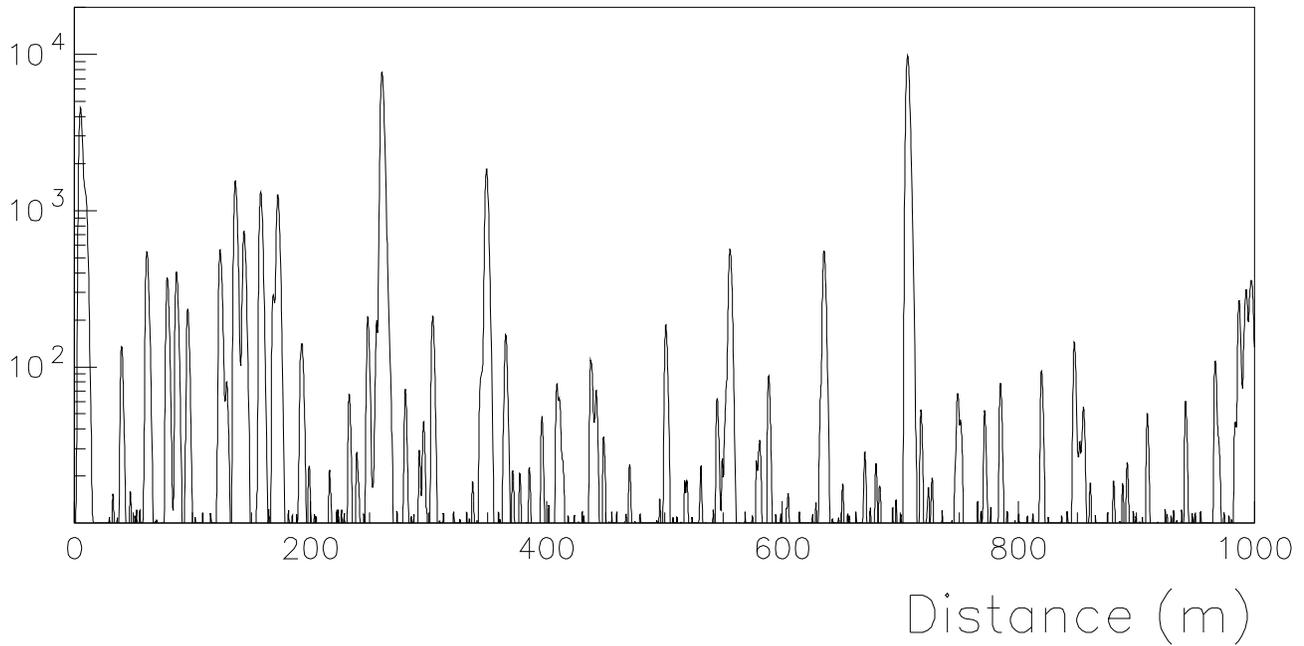
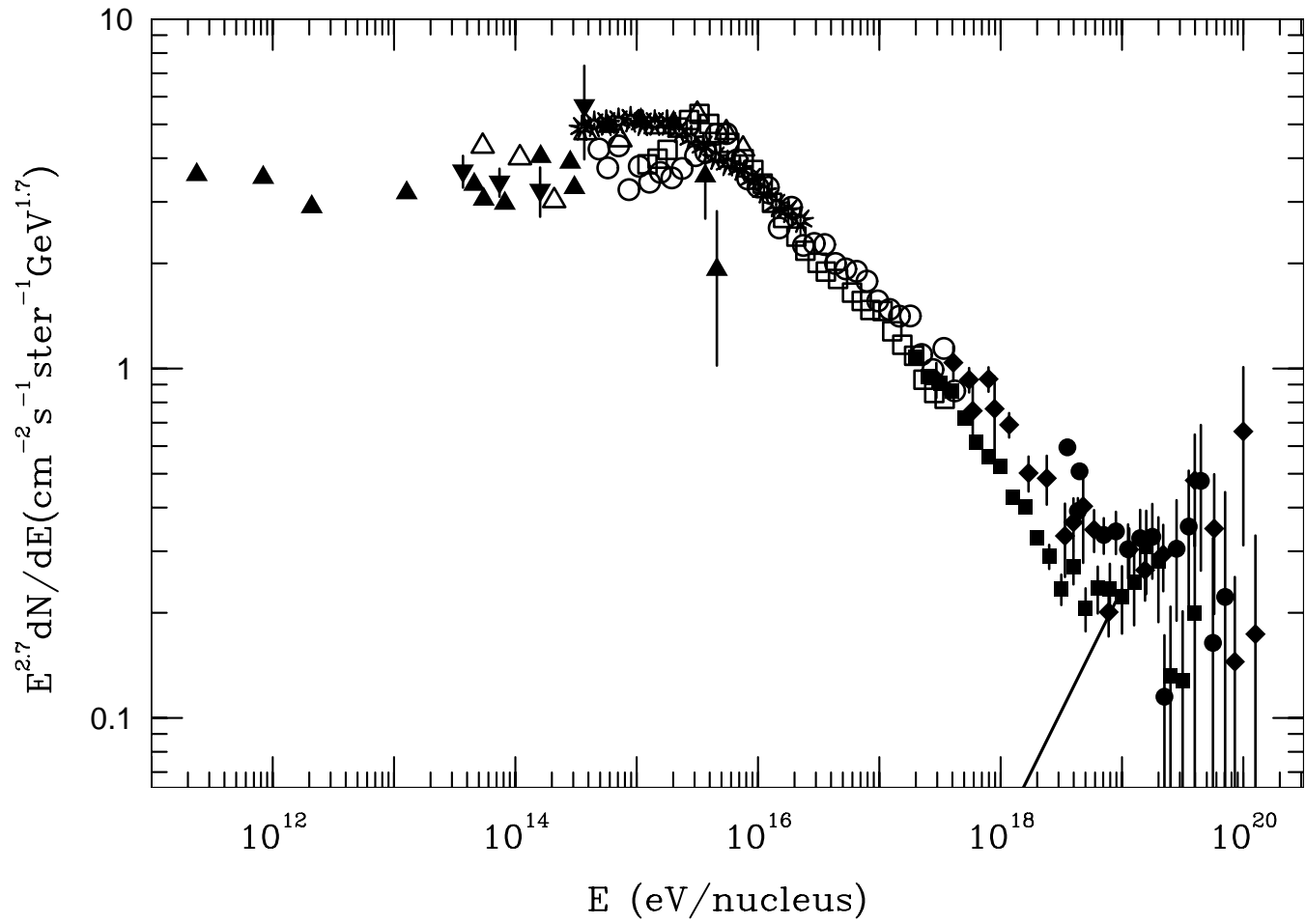
\end{document}